\documentclass[letterpaper, 10 pt, conference]{ieeeconf}  

\IEEEoverridecommandlockouts                              
\newcommand{\ep}{\mathcal{S}}

\newcommand{\cp}{\mathcal{R}}
\usepackage{cite}\usepackage{hyperref}
\usepackage{amsmath,amssymb,amsfonts}
\usepackage{algorithmic}
\usepackage{graphicx}
\usepackage{textcomp}
\usepackage{amsmath,amssymb,bm,bbm,mathrsfs,amscd}
\usepackage{calc}
\usepackage{color}
\usepackage{dsfont}
\usepackage{graphicx}
\usepackage{epstopdf}
\usepackage{epsfig}
\usepackage{tikz}
\usepackage{amsmath}
\usepackage{amsfonts}
\usepackage{amssymb}
\usepackage{rotating}
\usepackage{mathtools}
\usepackage{color}
\usepackage{subfig}
\usepackage{enumerate,pgfplots}
\newtheorem{theorem}{Theorem}
\newtheorem{definition}{Definition}

\newtheorem{lemma}{Lemma}

\newtheorem{example}{Example}

\newcommand{\ba}{\begin{array}}
	\newcommand{\ea}{\end{array}}

\newcommand{\be}{\begin{equation}}
\newcommand{\ee}{\end{equation}}

\newcommand{\mc}{\mathcal}

\def\1{\mathds{1}}
\def\0{\boldsymbol{0}}

\newcommand{\R}{\mathbb{R}}

\newcommand{\V}{\mathcal{V}}
\newcommand{\A}{\mathcal{A}}

\newcommand{\X}{\mathcal{X}}

\DeclareMathOperator*{\argmax}{argmax}

\def\R{\mathbb{R}}

\tikzstyle{v_c}=[circle, draw,inner sep=2pt, minimum width=12pt, color=blue]
\tikzstyle{v_a}=[circle, draw,inner sep=2pt, minimum width=12pt, color=red]
\tikzstyle{edge} = [draw,thick,-,font=\small ]
\tikzstyle{label} = [draw,fill=black,font=\normalsize]

\def\BibTeX{{\rm B\kern-.05em{\sc i\kern-.025em b}\kern-.08em
		T\kern-.1667em\lower.7ex\hbox{E}\kern-.125emX}}
\title{\LARGE \bf
Equilibria and learning dynamics in mixed network coordination/anti-coordination games
}

\author{Laura~Arditti, 
	Giacomo~Como,~\IEEEmembership{Member,~IEEE,}
	Fabio~Fagnani, and Martina Vanelli
	\thanks{The authors are with the  Department of Mathematical Sciences ``G.L.~Lagrange,'' Politecnico di Torino, 10129 Torino, Italy  (e-mail: {\{laura.arditti;\,giacomo.como;\,fabio.fagnani;\,martina.vanelli\}@polito.it}). G. Como is also with the Department of Automatic Control, Lund University, 22100 Lund, Sweden.}
	\thanks{This work was partially supported by a MIUR grant ``Dipartimenti di Eccellenza 2018--2022'' [CUP: E11G18000350001], a MIUR  Research Project PRIN 2017 ``Advanced Network Control of Future Smart Grids'' (http://vectors.dieti.unina.it) and by the Compagnia di San Paolo.}%
}

\begin{document}

	\maketitle
	\thispagestyle{empty}
	\pagestyle{empty}
	
	\begin{abstract}	
		Whilst network coordination games and network anti-coordination games have received a considerable amount of attention in the literature, network games with coexisting coordinating and anti-coordinating players are known to exhibit more complex behaviors. In fact, depending on the network structure, such games may even fail to have pure-strategy Nash equilibria. An example is represented by the well-known matching pennies (discoordination) game. 
		
		In this work, we first provide graph-theoretic conditions for the existence of pure-strategy Nash equilibria in mixed network coordination/anti-coordination games of arbitrary size. For the case where such conditions are met, we then study the asymptotic behavior of best-response dynamics and provide sufficient conditions for finite-time convergence to the set of Nash equilibria. Our results build on an extension and refinement of the notion of network cohesiveness and on the formulation of the new concept of network indecomposibility. 
		
	\end{abstract}

	
	\section{Introduction}
	%
	%
	%
	Coordination and anti-coordination games are two representative popular models of network games \cite{Jackson.Zenou:2015}, \cite{Ramazi.Riehl.Cao:2016} with a variety of applications in economics, social sciences, and biology. In their simplest version, they are mathematically described as strategic games with binary action set, where players are interconnected through a network. Specifically, the utility of a player in a network coordination (anti-coordination) game is an affine increasing (decreasing) function of the number of her neighbors in the network playing the same action. Despite their apparent similarity, both fundamental properties and applications of network coordination and network anti-coordination games are quite different. 	
	
        Network coordination games model the so called strategic complements effects, that is when the choice of a certain action by one player makes it more appealing for other players to play the same action. They are used to model social network features like the adoption of beliefs or behavioral attitudes, or economic ones such as the spread of a new technology. Mathematically, they belong to the broader class of supermodular games. As a consequence, Nash equilibria always exist and one special instance of them are the consensus configurations, namely, those where all individuals are playing the same action. 
	
	In contrast, network anti-coordination games are representative of another class exhibiting the so called strategic substitutes effect \cite{Bramoulle:2007}. In this case, the choice of a certain action by one player makes it more appealing for the other players to play the opposite action. They provide a natural model in situations where players are competing for resources that can become congested or in models where players can provide a public good, buy snob goods, or, in general, when there are gains from differentiation. While for strategic substitutes games, there is not a general mathematical theory as in the previous case, for the specific case of binary anti-coordination games Nash equilibria do exist but, differently from the coordination games, their form is strongly dependent on the network structure. 
	
	In this paper, we consider network games comprising both coordinating and anti-coordinating players. This class of games can be used to model the presence of anti-conformist behaviors in a social community, accounting for some form of  heterogeneity. More generally, games exhibiting both strategic complements and substitutes have been recently proposed in the economic literature \cite{Jackson.Zenou:2015}, \cite{Monaco:2016} to model heterogeneous interactions, e.g., markets with coexistence of both Cournot and Bertrand type firms. Such mixed games may fail to possess Nash equilibria. A fundamental example is the matching pennies game, which is a two-player game with one coordinating and the one anti-coordinating player.
	
	The focus of this paper is on the existence and the structure of pure strategy Nash equilibria in games with coordinating and anti-coordinating players as well as on the analysis of the best response dynamics in such games. Our contribution is twofold. First, we present a sufficient condition for the existence of pure strategy Nash equilibria whose restriction to the set of coordinating players is a consensus configuration. Second, we analyze conditions guaranteeing that learning rules like the best response dynamics converge to such equilibria Our results the relevance of the topological structure of the network as well as of the position of the coordinating and the anti-coordinating players within it. 
	
	The proposed sufficient conditions are of geometric type and build on the notion of cohesiveness introduced in \cite{Morris:2000} to describe the pure Nash equilibria of network coordination games. Our existence result requires the subset of coordinating players within the network to be cohesive itself. On the other hand, conditions for convergence of the best response dynamics are characterized in terms of a novel notion of indecomposability, related to the uniform non-cohesiveness property used in \cite{Morris:2000} to ensure full contagion in related threshold dynamical models. Our results may be interpreted in terms of robustness of pure network coordination games against the change of behavior of a subset of players: this viewpoint is developed in our recent work \cite{Arditti.ea:2021}. 
	
	Our first result is Theorem \ref{coro:mixed-existence} stating that a configuration with all coordinating players playing the same action can be an equilibrium for the whole game, despite the presence of anti-coordinating players, provided that the set of coordinating players is sufficiently cohesive in the network. 
	Our second result is Theorem \ref{coro:mixed-convergence}, establishing sufficient conditions for finite-time global convergence to such equilibria, showing that an action can spread to the whole population of coordinating players and be stable even in presence of anti-coordinating players. This result extends the analysis of contagion performed in \cite{Morris:2000} for pure coordination games to the mixed case.	
	These results are then extended in Theorem \ref{th:existence-extended} and Theorem \ref{th:convergence-extended} to the case of heterogeneous thresholds. 

	
	%
	%
	%
	%
	%
	%
	%
	%
	
	\subsection{Related work}
	Pure network coordination and anti-coordination games have been extensively studied in the literature.  We refer to \cite{Jackson.Zenou:2015,Ramazi.Riehl.Cao:2016,Bramoulle:2007,Montanari.Saberi:2010,Jackson.Storms:2019,Eksin.Paarporn:2020,Como.Durand.Fagnani:2020} and references therein for the most recent results. In this work we will make use of the results in \cite{Morris:2000} on the Nash equilibria of coordinating game, particularly as presented and discussed in \cite{Jackson.Zenou:2015}.  
	Mixed games with the simultaneous presence of coordinating and anti-coordinating players have received limited attention. They are introduced in \cite{Ramazi.Riehl.Cao:2016} but no theoretical analysis is provided. 
	
	Somehow related is the linear threshold model \cite{Granovetter:1978,Lelarge:2012,Rossi.ea:2017}. This is a dynamical system where, synchronously, all players update their action choosing the one adopted by the majority in their neighborhood. This model has been extended \cite{Grabisch.ea:2019,Juul.Porter:2019,Nowak.ea:2019} considering also the possible presence of anticonformist individuals that choose to play the action played by the minority in their neighborhood. Various versions of the linear threshold model have been introduced depending on how matches are treated and depending if certain transitions are considered or not to be irreversible. In any case, absorbing states of such dynamical systems are Nash equilibria of the corresponding game and its asymptotics is related to the asymptotics of other learning rules. Results are however restricted to the case of a complete graph or of a random graph (where neighborhood is changed at every instant). 
	
	The game considered in this paper was studied in \cite{Vanelli.ea:2020} in the spacial case of a complete graph. Preliminary results for general graphs have appeared in the MS thesis \cite{Vanelli:2019}. 

	\section{Problem formulation}\label{sec:problem-formulation}
	In this section, we introduce the mixed network coordination/anti-coordination game and present the main issues addressed in the rest of the paper.
	\subsection{Definition of the game}
	Let $\mc G=(\mc V, \mc E, W)$ be a finite undirected weighted graph with node set  $\mc V$, link set $\mc E$, and nonnegative symmetric weight matrix $W=W'$ in $\R_+^{\mc V\times\mc V}$ such that $W_{ij}=W_{ji}>0$ if and only if $\{i,j\}$ is a link in $\mc E$. We shall not allow for the presence of self-loops, so that $W_{ii}=0$ for all $i$ in $\mc V$. 
	For a subset of nodes $\mc U\subseteq\mc V$, we shall denote by  
	$$w_i^{\mc U} = \sum_{j\in \mc U} W_{ij}\,,\qquad i\in\mc V\,,$$
	the $\mc U$-\textit{restricted degree} of the nodes. 
	In the special case $\mc U=\mc V$, we simply refer to $w_i=w_i^{\mc V}$ as the \textit{degree} of node $i$.

	We  shall consider strategic form games whose players are identified with the nodes of $\mc G$ and where each player $i$ in $\mc V$ chooses a binary action $x_i$ in $\A=\{0,1\}$ aiming to maximize her utility  
	\begin{equation}\label{eq:ut}
	u_i(x_i,x_{-i})=				\delta_i\sum_{j\in\V}W_{ij}\big((1-r)x_ix_j + r(1-x_i)(1-x_j)\big)\,.
	\end{equation}
	In \eqref{eq:ut}, $x_{-i}$ in $\mc A^{\mc V\setminus\{i\}}$ stands for the strategy profile of all players except for $i$,  $r$ in $(0,1)$ is a common threshold value, while $\delta_i$ in $\{\pm1\}$ charaterizes player $i$'s behavior as coordinating or anti-coordinating. Specifically, we refer to 
	$$\mc R=\{i\in\mc V:\,\delta_i=+1\}\,,\qquad\mc S=\{i\in\mc V:\,\delta_i=-1\}\,,$$
	as the sets of coordinating and anti-coordinating players, respectively. 
	Notice that the utility of a coordinating player $i\in\cp$ 
	increases with every neighbor playing the same action, while the utility of an anti-coordinating player $i\in \ep$ 
	decreases with every neighbor playing the same action. More precisely, a coordinating (resp. anti-coordinating) player  assigns a weight $W_{ij}(1-r)$ (resp $-W_{ij}(1-r)$) to a match on action $1$ with player $j$, while a match on action $0$ with the same player weights  $W_{ij}r$ (resp. $-W_{ij}r$). 
	
\begin{figure}
		\centering
		\scalebox{0.8}{%
		\begin{tikzpicture}[scale=1.1]
		\foreach \x/\name in {(1.3,0)/1, (2,-1)/2, (2.7,0)/3, (4,-1)/4, (4,0)/5,(5,-1)/6, (2,1)/7, (3.5,1)/8/-1}\node[shape=circle,draw=blue,text=blue](\name) at \x {\small\name};
		\foreach \x/\name in {(6,0)/9,(3,-1.5)/10,(1,-1.5)/11,(0,0)/12,(5,1)/13}\node[shape=circle,draw=red,text=red](\name) at \x {\small\name};
		
		
		\foreach \a/\b/\w in {1/2/2,2/3/1, 4/6/2,7/8/3,5/9/4, 2/10/2,2/11/1, 1/12/4,8/13/1, 13/9/5,3/13/7}\path [-,draw] (\a) edge node[above] {\small\w} (\b);
		
		\foreach \a/\b/\w in {3/5/3,  5/13/2,  11/10/1, 4/10/2,3/4/1}\path [-,draw] (\a) edge node[below] {\small\w} (\b);
		
		\foreach \a/\b/\w in {7/3/1, 4/5/2,12/11/2}\path [-,draw] (\a) edge node[right] {\small\w} (\b);
		
		\foreach \a/\b/\w in {1/7/3}\path [-,draw] (\a) edge node[left] {\small\w} (\b);

		\end{tikzpicture}
	}
		\caption{Undirected weighted graph $\mc G$ where weights are represented by the values on the links. The set of coordinating players $\cp=\{1,\dots,8\}$ is depicted in blue, while the set of anti-coordinating ones $\ep=\{9,\dots,13\}$ is red.}
		\label{fig:perturbedcoordination}
	\end{figure}
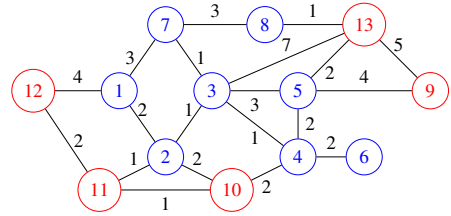

	The setting is depicted in Figure \ref{fig:perturbedcoordination}. Throughout the paper, we will refer to this game as \textit{mixed network coordination/anti-coordination game}, often using the acronym CAC. In the special cases when only one type of players is present, namely $\cp = \V$ or $\ep=\V$, we simply call it  
	\textit{network coordination game} or, respectively, \textit{network anti-coordination game}.

	The \textit{best response} (BR) correspondence for player $i \in \mc V$ gives the set of the best actions for $i$ given the configuration of other players. Formally,
	$$\mc B_i(x_{-i})= \argmax_{x_i \in\mc A}u_i(x_i,x_{-i}) \,.$$

	The best response of a coordinating player $i\in \cp$ is $1$ when the total weight of players choosing action $1$ is \emph{above} a fraction $r$ of the total degree, while the best response of an anti-coordinating player $i\in \ep$ is $1$ when the total weight of players choosing action $1$ is \emph{below} a fraction $r$ of the total degree. 
	This consideration clarifies the meaning of the parameter $r$, which determines the behavior of a player by specifying the threshold in the fraction of her neighbors playing action $1$ at which her preference for action $1$ changes.

	The purpose of this work is to investigate mixed network  
	CAC games for existence and reachability of (pure strategy) \emph{Nash equilibria} i.e., configurations $x^*$ in $\mc X$ such that 
	$$x^*_i\in\mc B_i(x^*_{-i})\,,\qquad\forall i\in\mc V\,.$$
	The set of equilibria of a game will be denoted as $\mc N$. 
	
	The last part of this section is devoted to clarify the concept of reachability of the set of Nash equilibria. In our setting, we consider a dynamics where only one player  at a time modifies its action. More precisely, two consecutive configurations differ just for the action of a single player, who modifies her action according to the best response function.

	Formally, consider two strategy profiles $x,y\in \X$. For $l\ge0$, a length-$l$ \emph{best response path} (\emph{BR-path}) from $x$ to $y$ is an $(l+1)$-array of strategy profiles in $\mc X^{l+1}$, denoted with 
	$(x^{(0)},x^{(1)},\ldots x^{(l)})$,  such that 
	\begin{itemize}
		\item $x^{(0)}=x$, and $x^{(l)}=y$; 
		\item for every $k=1,2,\ldots, l$, there exists a player $i_k$ in $\mc V$ such that  
		\be\label{eq:BRpath} x^{(k)}_{-i_k}=x^{(k-1)}_{-i_k}\,, \qquad 	x_{i_k}^{(k)}\in\mc B_{i_k}(x_{-i_k}^{(k-1)})\,.\ee 
	\end{itemize}
	We shall refer to a subset of strategy profiles $\mc Y\subseteq\mc X$ as:
	\begin{itemize}
		\item \emph{reachable} from strategy $x$ in $\mc X$ if there exists a BR-path from $x$ to some strategy profile $y$ in $\mc Y$;  
		\item \emph{globally} reachable if it is reachable from every configuration $x$ in $\mc X$.
	\end{itemize}
	
	To indicate that $y$ is reachable from $x$, we will use the notation $x\to y$.

	\subsection{Some examples}

	We end this section with a number of simple examples illustrating how the aforementioned properties can fail to show up in such games. We start with the notable matching pennies game.
	
	\begin{example}
		Consider a simple graph with two nodes connected by an undirected link, where the first node is coordinating and the other one is anti-coordinating. 
		When $r=\frac{1}{2}$, this reduces to the discoordination game which is well-known not to admit pure strategy Nash equilibria. 
	\end{example}
	
	
	The previous example is a  special case of the mixed CAC game defined on a complete graph, i.e., $W_{ij}=1$ for all $i,j\in \V$, $i\neq j$. A detailed study of existence and characterization of Nash equilibria in mixed CAC games on the complete graph with heterogeneous thresholds $r_i$ for every player $i\in \mc V$ was proposed in \cite{Vanelli.ea:2020}. 
	
\begin{figure}
		\centering
		\scalebox{0.8}{%
		\begin{tikzpicture}[scale = 0.5]
		\node[shape=circle,draw=red, text=red] (A) at (0.5,0) {\small1};
		\node[shape=circle,draw=blue, text=blue] (B) at (0,1.5) {\small2};
		\node[shape=circle,draw=blue, text=blue] (C) at (1.5,2.5) {\small3};
		\node[shape=circle,draw=blue, text=blue] (D) at (2.5,0) {\small5};
		\node[shape=circle,draw=blue,text=blue] (E) at (3,1.5) {\small4};
		\node[shape=circle,draw=blue,text=blue] (F) at (-1,0) {\small6};
		\node[] () at (-1,2.5) {A.};
		
		\foreach \a in {D,E}
		\foreach \b in {C,D}
		\path [-] (\a) edge node {} (\b);
		\path [-,draw] (B) edge node {} (C);
		\path [-,draw] (A) edge node {} (B);
		\path [-,draw] (A) edge node {} (F);
		\path [-,draw] (A) edge node {} (D);	
		\path [-,draw] (F) edge node {} (B);	
		
		\end{tikzpicture}\hspace{0.2cm}
		\begin{tikzpicture}[scale = 0.5]
		
		\node[shape=circle,draw=red, text=red] (A) at (0.5,0) {\small1};
		\node[shape=circle,draw=blue, text=blue] (B) at (0,1.5) {\small2};
		\node[shape=circle,draw=blue, text=blue] (C) at (1.5,2.5) {\small3};
		\node[shape=circle,draw=blue, text=blue] (D) at (2.5,0) {\small5};
		\node[shape=circle,draw=blue,text=blue] (E) at (3,1.5) {\small4};
		\node[shape=circle,draw=blue,text=blue] (F) at (-1,0) {\small6};
		\node[] () at (-1,2.5) {B.};
		
		\foreach \a in {D,E}
		\foreach \b in {C,D}
		\path [-] (\a) edge node {} (\b);
		\path [-,draw] (B) edge node {} (C);
		\path [-,draw] (A) edge node {} (B);
		\path [-,draw] (A) edge node {} (F);
		\path [-,draw] (A) edge node {} (D);
		
		\end{tikzpicture}\hspace{0.2cm}
		\begin{tikzpicture}[scale = 0.5]
		
		\node[shape=circle,draw=red, text=red] (A) at (0.2,0) {\small1};
		\node[shape=circle,draw=blue, text=blue] (B) at (-0.3,1.5) {\small2};
		\node[shape=circle,draw=blue, text=blue] (C) at (1.2,2.5) {\small3};
		\node[shape=circle,draw=blue, text=blue] (D) at (2.2,0) {\small5};
		\node[shape=circle,draw=blue,text=blue] (E) at (2.7,1.5) {\small4};
		\node[shape=circle,draw=blue,text=blue] (F) at (-1.3,0) {\small6};
		\node[] () at (-1.3,2.5) {C.};
		
		\foreach \a in {D,E}
		\foreach \b in {C,D}
		\path [-] (\a) edge node {} (\b);
		
		\path [-,draw] (A) edge node {} (B);
		\path [-,draw] (A) edge node {} (F);
		\path [-,draw] (A) edge node {} (D);
		
		\end{tikzpicture}
	}
		\caption{The three graphs considered in Example \ref{ex:network}, with  $\cp=\{2,\dots, 6\}$ (in blue) and $\ep=\{1\}$ (in red). When $r=\frac{1}{2}$, a Nash equilibrium exists for both the graphs A and B, while it does not for graph C.} 
		\label{fig:network}
	\end{figure}
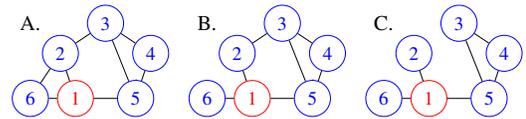

	The network structure makes the problem more complex, making the interconnections among players play a crucial role. 
	\begin{example}\label{ex:network}
		Let us consider the mixed network CAC game with threshold $r=\frac{1}{2}$ on the three graphs in Figure \ref{fig:network}. When players interact according to the graph A, the game admits two Nash equilibria where all coordinating players coordinate on one action while the anti-coordinating player picks the opposite one: $x^*$  and $\1-x^*$, with $x^*_i = 1$ for $i\in \cp$ and $x^*_i=0$ for $i\in\ep$. If we remove the edge between coordinating players $2$ and $6$, as in graph B, the two configurations $x^{**}$ and $\1-x^{**}$ with $x^{**}_i = 1$ for $i \in \{2,3,4,5\}$ and $x^{**}_i = 0$ for $i\in \{1,6\}$ become Nash equilibria of the game. If we remove also the edge between player $2$ and $3$, obtaining graph C, the game no longer admits Nash equilibria.
	\end{example}
	
	In general, existence of Nash equilibria depends on the network structure, the value of the threshold parameter, the roles of players, and on the relationship among these three features.  
	Furthermore, even when the set of Nash equilibria is non-empty, it might not be globally reachable, as shown by the following example.

\begin{figure}
		\centering
		\vspace{0.2cm}
	\scalebox{0.8}{%
		\begin{tikzpicture}[scale=0.8]
		\foreach \x/\name in {(1,0)/1, (3,0)/3, (2,-1)/4, (2,1)/2, (4,0)/5, (6,0)/7, (5,-1)/8, (5,1)/6,(3.5,-1)/9}\node[shape=circle,draw=blue, text=blue](\name) at \x {\small \name};
		\node[shape=circle,draw=red,text=red](10) at (3.5,-2) {10};
		\path [-] (9) edge node {} (10);
		\foreach \i in {8,4}{	
			\path [-,draw] (\i) edge node {} (9);
			\path [-,draw] (\i) edge node {} (10);}	
		\foreach \i in {5,6,7,8}
		{\foreach \j in {5,6,7,8}
			\path [-,draw] (\i) edge node {} (\j);}
		\foreach \i in {1,2,3,4}
		{\foreach \j in {1,2,3,4}
			\path [-,draw] (\i) edge node {} (\j);}	
		\end{tikzpicture}
	}
		\caption{The graph considered in Example \ref{ex:no-conv}, with $\cp = \{1, \dots, 9\}$ (in blue) and $\ep = \{10\}$ (in red). When $r=\frac{1}{2}$, the mixed network CAC game admits at least two Nash equilibria, but the set $\mc N$ is not globally reachable. 
		}
		\label{fig:no-conv}
	\end{figure}
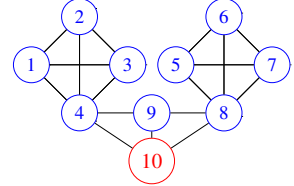

	\begin{example}\label{ex:no-conv}
		Consider a mixed network CAC game over the graph in Figure \ref{fig:no-conv} with $r=\frac{1}{2}$. The game admits the two Nash equilibria $x^*$ and $\1-x^*$, with $x^*_i=1$ for $i\in \cp$ and $x^*_i=0$ for $i\in \ep$. Anyway, in general, convergence to a Nash equilibrium is not guaranteed. Notice that, starting with any initial condition $x$ satisfying 
		$x_1=x_2=x_3=x_4=1$ and $x_5=x_6=x_7=x_8=0$,
		we will never reach a Nash equilibrium.
	\end{example}
	The study of Nash equilibria for the mixed network CAC game is then a challenging problem. 
	In the rest of this work, we will investigate the existence and reachability of Nash equilibria in this complex and interesting setting.
	
	\section{Main results}\label{sec:mixed-games}
	
	In this section we will explore how the network properties reflect on both static and dynamic properties of network CAC games, by analysing how the presence of anti-coordinating players in $\ep$ and the structure of interconnections affects the behavior of the coordinating players in $\cp$. The outline of the proofs of Theorem \ref{coro:mixed-existence} and \ref{coro:mixed-convergence} can be found in Section \ref{s:proofs}.
	
	In what follows we focus on Nash equilibria that are consensus in the $\cp$ side. 
	First we investigate the existence of such equilibria obtaining Theorem \ref{coro:mixed-existence}, which shows how the structure of interactions is determinant in the mixed game. In particular, the cohesiveness property of $\mc G$ will play a crucial role.
	
	
	\begin{definition}[cohesiveness]\label{def:cohesiveness}
		Given a graph $\mc G=(\mc V,\mc E,W)$ and a threshold $r\in  (0,1)$, a subset of nodes $\mc S\subseteq\mc V$ is called $r$-\emph{cohesive} in $\mc G$ if 
		\be\label{eq:cohesiveness}w_i^{\mc S}\ge rw_i\,,\qquad\forall i\in\mc S\,.\ee
	\end{definition}

	\begin{theorem}\label{coro:mixed-existence}
		Consider a network CAC game on an undirected graph $\mc G=(\mc V,\mc E, W)$, with set of coordinating and anti-coordinating players $\cp\subseteq\mc V$ and $\ep=\mc V\setminus\cp$  respectively and threshold $r$.
		\begin{enumerate}
			\item If $\cp$ is $r$-cohesive then 
			there exists at least one Nash equilibrium where all coordinating players play action $1$. 
			\item  If $\cp$ is $(1-r)$-cohesive then 
			there exists at least one Nash equilibrium where all coordinating players play action $0$.
		\end{enumerate}
	\end{theorem}
	
	Theorem \ref{coro:mixed-existence} provides a sufficient graph-theoretic condition for the existence of pure strategy Nash equilibria, namely, cohesiveness of the set of coordinating players, and it significantly generalizes previous works where existence of pure strategy Nash equilibria was proved only for pure coordination or anti-coordination games \cite{Ramazi.Riehl.Cao:2016}. 
	
	
	\begin{example}\label{ex:mixed-existence}
		Consider a mixed network CAC game with $r=0.4$ over the graph $\mc G$ in Figure \ref{fig:perturbedcoordination}, where nodes in $\cp =\{1, \ldots, 8\}$ (in blue) are coordinating players, while nodes in $\ep =\mc V \setminus \cp$ (in red) are anti-coordinating players. Since $\cp$ is $0.4$-cohesive in $\mc G$, the game admits at least one Nash equilibrium $x^*$
		where $x^*_i=0$ for all $i\in \cp$. For instance, if
		$x^*_9= x^*_{11}=0$ and $x^*_{10}= x^*_{12}= x^*_{13}=1$ then $x^*$ is Nash equilibrium of the game 
	\end{example}
	We remark that the condition is sufficient but not necessary: a Nash equilibrium can exist even if the subset of the coordinating players is not $r$-cohesive. 
	
	\begin{example}\label{ex:non-suff}
		Consider the mixed network CAC game on the graph B of Figure \ref{fig:network} with $r=\frac{1}{2}$. 
		Even if the $\frac{1}{2}$-cohesiveness property is violated by player $6$, a Nash equilibrium does exist as observed in Example \ref{ex:network}. 
	\end{example}
	
	
	Theorem \ref{coro:mixed-existence} guarantees existence of Nash equilibria but not uniqueness. In general, there might be other Nash equilibria where the coordinating players' configuration is different from a consensus.

	We now focus on the reachability of the set
	of Nash equilibria for which we proved existence in Theorem \ref{coro:mixed-existence}, that is, those where the population of coordinating players is at consensus. 
	
	To address this point, we need to introduce one further geometric notion for the graph $\mc G$.
	
	\begin{definition}[indecomposability]\label{definition:robust_non_cohesiveness}
		Consider a graph $\mc G=(\mc V,\mc E, W)$, a subset of nodes $\cp\subseteq\mc V$ and its complementary $\ep= \mc V \setminus \cp $ and a threshold $r\in (0,1)$. We say that $\cp$ is $r$-\emph{indecomposable} in $\mc G$ if for every partition $\cp = \cp_0 \cup \cp_1$, $\exists i \in \cp$ such that either
		\begin{equation}\label{eq:indec1}
		i \in \cp_1 \text{ and } w_i^{\cp_1} + w_i^{\ep} < r w_i
		\end{equation}
		or
		\begin{equation}\label{eq:indec0}
		i \in \cp_0 \text{ and } w_i^{\cp_0} + w_i^{\ep} < (1-r) w_i \,.
		\end{equation}
	\end{definition}
	
	While cohesiveness is a well-known concept, first introduced in \cite{Morris:2000}, indecomposability is a new geometric notion that extends the \textit{uniform no more than $\frac{1}{2}$-cohesiveness} property proposed in \cite{Morris:2000} to the case where anti-coordinating players and a generic threshold $r$ are present. 
	It requires that for each partition of $\cp$ into two non-empty sets $\cp_0$ and $\cp_1$ at least one of the two sets $\cp_1 \cup \ep$ or $\cp_0 \cup \ep$ is not sufficiently cohesive, where cohesiveness is violated by nodes in $\cp_1$ or $\cp_0$ respectively. 
	
\begin{figure}
		\centering
		\scalebox{0.8}{%
		\begin{tikzpicture}[scale=0.7]
		\draw (0.4,1) circle (50pt);
		\node () at (0.4,3.2) {$\large\mc R_0$};
		\draw (4.8,1) circle (50pt);
		\node () at (4.8,3.2) {$\large\mc R_1$};
		\node[shape=circle,draw=blue,text=blue] (A) at (0,0) {1};
		\node[shape=circle,draw=blue,text=blue] (B) at (0,2) {2};
		\node[shape=circle,draw=blue,text=blue] (C) at (1.5,1) {3};
		\node[shape=circle,draw=blue,text=blue] (D) at (5,2) {5};
		\node[shape=circle,draw=blue,text=blue] (E) at (5,0) {6};
		\node[shape=circle,draw=blue,text=blue] (F) at (4,1) {4} ;
		\node[shape=circle,draw=red,text=red](G) at (-2,0) {7};
		\node[shape=circle,draw=red,text=red](H) at (-2,2) {8};
		\node[shape=circle,draw=red,text=red](I) at (-2,1) {9};
		\node[shape=circle,draw=red,text=red](L) at (7,2) {\small 10};
		\node[shape=circle,draw=red,text=red](M) at (2.5,2.2) {\small11};
		\node[shape=circle,draw=red,,text=red] (N) at (7,0) {\small 12};
		
		\foreach \a/\b in {A/B, A/C, B/C, D/C, A/E, D/E, D/F, C/F, E/F}\path [-,draw](\a) edge node {} (\b);
		\foreach \a/\b in {A/G, B/H, A/I, D/L, C/M, E/N}\path [-,draw](\a) edge node {} (\b);
		\end{tikzpicture}
	}
		\caption{The graph considered in Example \ref{ex:cond-not-sat}. The set $\cp=\{1,\dots,6\}$ (in blue) is not $\frac{1}{2}$-indecomposable.}
		\label{fig:cond-not-sat}
	\end{figure}
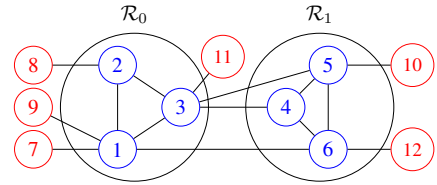

	\begin{example}\label{ex:cond-not-sat}
		Consider the graph $\mc G$ in Figure \ref{fig:cond-not-sat}.
		The set $\cp$ in blue is \textit{not} $\frac{1}{2}$-indecomposable as there exists a partition of $\mc R$ into $\mc R_0=\{1,2,3\}$ and $\mc R_1=\{4,5,6\}$ such that both $\mc R_0\cup \ep$ and $\mc R_1\cup \ep$ are $\frac{1}{2}$-cohesive in $\mc G$. 
	\end{example}

	In the following result Theorem \ref{coro:mixed-convergence}, we show that indecomposability 
	combined with the cohesiveness of $\cp$ guarantees that the set of Nash equilibria where coordinating players are at consensus is globally reachable.
	
	\begin{theorem}\label{coro:mixed-convergence}
		Consider a network CAC game on an undirected graph $\mc G=(\mc V,\mc E, W)$, with set of coordinating and anti-coordinating players $\cp\subseteq\mc V$ and $\ep=\mc V\setminus\cp$  respectively and threshold $r$.
		Assume that
		\begin{enumerate}
			\item[a)] $\cp$ is $r$-cohesive or $(1-r)$-cohesive,
			\item[b)] $\cp$ is $r$-indecomposable in $\mc G$.
		\end{enumerate}
		Then 
		the set of Nash equilibria where coordinating players are at consensus is non-empty and it is globally reachable.
	\end{theorem}

\begin{figure}
		\centering
		\vspace{0.2cm}
		\scalebox{0.8}{%
		\begin{tikzpicture}[scale = 0.65]
		
		\node[shape=circle,draw=blue, text=blue] (A) at (0,0) {1};
		\node[shape=circle,draw=blue, text=blue] (B) at (0,1.5) {2};
		\node[shape=circle,draw=blue, text=blue] (C) at (1.5,2.5) {3};
		\node[shape=circle,draw=blue, text=blue] (D) at (3,0) {5};
		\node[shape=circle,draw=blue, text=blue] (E) at (1.5,-1) {6};
		\node[shape=circle,draw=blue, text=blue] (F) at (3,1.5) {4} ;
		
		\node[shape=circle,draw=red,text=red] (G) at (-2,0) {7};
		\node[shape=circle,draw=red,text=red] (H) at (-2,1.5) {8};
		\node[shape=circle,draw=red,text=red] (L) at (5,0) {9};
		\node[shape=circle,draw=red,text=red] (M) at (4,2.5) {\small 10};
		\node[shape=circle,draw=red,text=red] (N) at (3.5,-1) {\small 11};
		
		\foreach \a in {A,B,C,D,E,F}
		\foreach \b in {A,B,C,D,E}
		\path [-,draw] (\a) edge node {} (\b);
		
		\foreach \a/\b in {A/G, B/H, F/L, D/L, A/G, D/L, C/M, E/N, L/M, L/N, G/H}
		\path [-, draw] (\a) edge node {} (\b);
		\end{tikzpicture}
	}
		\caption{The graph in Example \ref{ex:convergence} where $\cp=\{1,\dots, 6\}$ is blue and $\ep=\{7,\dots, 11\}$ is red. When $r=\frac{1}{2}$, both the hypothesis of Theorem \ref{coro:mixed-convergence} are satisfied and convergence to 
			a Nash equilibrium where coordinating players are at consensus 
			is guaranteed.  }
		\label{fig:cond-sat}
	\end{figure}
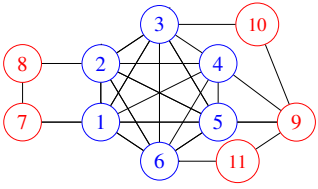

	Theorem \ref{coro:mixed-convergence} can be seen as the analogous of the analysis of the contagion process performed in \cite{Morris:2000} for the network coordination game, where we investigate how an action can spread to the whole coordinating population and be stable despite the presence of the anti-coordinating players.
	\begin{example}\label{ex:convergence}
		Consider a mixed network coordination-anti-coordination game with $r=\frac{1}{2}$ over the graph $\mc G$ in Figure \ref{fig:cond-sat}, where nodes in $\cp =\{1,\ldots,6\}$ (in blue) are coordinating players, while nodes in $\ep =\mc V \setminus \cp$ (in red) are anti-coordinating players. Since $\cp$ is $\frac{1}{2}$-cohesive in $\mc G$ the game admits at least two Nash equilibria $x^*$ and $\1-x^*$ where $x^*_i = 1$ for all $i \in \cp$. Furthermore, since $\cp$ is $\frac{1}{2}$-indecomposable, convergence to such equilibria is guaranteed from any initial condition.
	\end{example}
	\begin{example}
		Consider a mixed network coordination/anti-coordination game over the graph in Figure \ref{fig:no-conv} with $r=\frac{1}{2}$. The set $\cp=\{1,\ldots,8\}$ is $\frac{1}{2}$-cohesive, which implies that there exist at least two Nash equilibria, 
		where players in $\cp$ coordinate on action $1$ and $0$.
		Anyway, as $\cp$ is not $\frac{1}{2}$-indecomposable, convergence to the set of such equilibria is not guaranteed. Notice that, starting with any initial condition $x$ satisfying $x_1=x_2=x_3=x_4=1$ and $x_5=x_6=x_7=x_8=0$, we will never reach consensus of coordinating players. Actually, as observed in Example \ref{ex:no-conv}, from such initial condition, we will never converge to any Nash equilibrium.
	\end{example}
	We remark that the conditions in Theorem \ref{coro:mixed-convergence} are sufficient but not necessary. More precisely, the set of Nash equilibria where coordinating players are at consensus can be non-empty and globally reachable even if the $r$-indecomposability property is not satisfied, as shown in the following example.
	\begin{example}\label{ex:th2-suff}
		Let us consider a three-node complete graph $\mc G$ with $W_{ij} = 1$ if $i \neq j$, $i,j \in \mc V =\{1,2,3\}$. Let $r = \frac{1}{2}$, $\mc R=\{1,2\}$ and $\mc S = \{3\}$. Notice that $\mc R$ is $\frac{1}{2}$-cohesive but not $\frac{1}{2}$-indecomposable: the property is indeed violated by the sets $\mc R_0 = \{1\}$ and $\mc R_1 =\{2\}$. Anyway, the set of the two Nash equilibria $x^*$ and $\1-x^*$, with $x^*_3 = 1$ and $x^*_1=x^*_2 = 0$, is globally reachable.
\end{example}

	\section{Outline of proofs and extensions}\label{s:proofs}
	This section is devoted to the presentation of the outline of the proofs   
	of our results and to the description of some relevant properties of the coordination and anti-coordination game that play a significant role in our analysis.
	
	
	Actually, we will prove the results in the more general setting 
	where players may have heterogeneous thresholds collected in a vector $\mathbf{r}$. 
	More precisely, we shall consider the \textit{mixed network coordination/anti-coordination game} with 
	\textit{threshold vector} $\mathbf{r}= (r_i)_{i \in \mc V}$ with $r_i \in  (0,1)$, 
	which is defined in the exact same way of \eqref{eq:ut} 
	except for the replacement, for each player $i\in \V$, of the threshold $r$ with the $i$-th entry of the threshold vector, i.e., $r_i$.
	
	The notions of cohesiveness and indecomposability admit a natural extension to this more general setting.
	\begin{definition}
		Consider a graph $\mc G=(\mc V,\mc E, W)$, a subset of nodes $\cp\subseteq\mc V$ and its complementary $\ep= \mc V \setminus \cp $, and a vector of thresholds $\mathbf{r}= (r_i)_{i \in \mc V}$ with $r_i \in  (0,1)$. We say that $\cp$ is
		\begin{itemize}
			\item[(i)]$\mathbf{r}$-\emph{cohesive} in $\mc G$ if 
			\begin{equation}
			w_i^{\mc R}\ge r_iw_i\,,\qquad\forall i\in\mc R\,;
			\end{equation}
			\item[(ii)] $\mathbf{r}$-\emph{indecomposable} in $\mc G$ if for every partition $\cp = \cp_0 \cup \cp_1$, we have that $\exists i \in \cp$ such that either
			\begin{equation}
			i \in \cp_1 \text{ and } w_i^{\cp_1} + w_i^{\ep} < r_i w_i\,,
			\end{equation}
			or
			\begin{equation}
			i \in \cp_0 \text{ and } w_i^{\cp_0} + w_i^{\ep} < (1-r_i) w_i\,.
			\end{equation}
		\end{itemize} 	
	\end{definition}	
	Clearly, when $\mathbf{r} = r \1$ with $r \in (0,1)$ the previous notions reduce to Definition \ref{def:cohesiveness} and \ref{definition:robust_non_cohesiveness}.

To derive our results about the mixed game, it proves useful to represent the mixed network CAC game as a graphical game such that its restrictions to players $\cp$ and $\ep$ are network coordination and anti-coordination games, respectively. 
		To clarify the notion of restricted game, we first canonically identify the space of strategy profiles as the external product $\mc X=\mc X_{\cp}\times \mc X_{\ep}$ of the configuration spaces $\mc X_{\cp}=\mc A^{\cp}$ of players in $\cp$ and $\mc X_{\ep}=\mc A^{\ep}$ of players in $\ep$. We shall also decompose every configuration $x$ in $\mc X$ as $x=(y, z)$ with $y=\left.x\right|_{\cp}$ in $\mc X_{\cp}$ and $z=\left.x\right|_{\ep}$ in $\mc X_{\ep}$.
		For a strategy profile $z$ in $\mc X_{\ep}$ of the anti-coordinating players, we shall call \emph{$\cp$-restricted game} the game with player set $\cp$, configuration space $\mc X_\cp$ and utility functions $u_i^{(z)}(y)=u_i(y,z)$. Similarly, for a given $y$ in $\mc X_{\cp}$, the \emph{$\ep$-restricted game} has player set $\ep$, configuration space $\mc X_\ep$ and utility functions $u_i^{(y)}(z) = u_i(y,z)$.
	
	Such representation plays a crucial role in our analysis since, when $\cp$ and $\ep$ are the sets of coordinating and anti-coordinating players of a network CAC game, both the $\mc R$-restricted game and the $\mc S$-restricted game are potential games, i.e., games for which there exists a potential function $\Phi: \mc X \rightarrow \R$ such that \begin{equation}\label{potential_function}
	u_i(y_i, x_{-i})-u_i(x_i, x_{-i}) = \Phi(y_i, x_{-i})-\Phi(x_i, x_{-i}) 
	\end{equation}
	for all $x_i,y_i \in \A$, $x_{-i}\in \A^{\V\setminus \{i\}}$. Potential games \cite{Monderer.Shapley:1996} are a special class of games for which not only existence of Nash equilibria is guaranteed, but a non-empty subset of the set of Nash equilibria is globally reachable. The potentiality of the restricted games is proved in the following Lemma.
	\begin{lemma}\label{lemma1} Consider the mixed network CAC game with  threshold vector $\textbf{r}=(r_i)_{i\in \mc V}$,  on an undirected graph $\mc G=(\mc V,\mc E, W)$, with set of coordinating and anti-coordinating players $\cp\subseteq\mc V$ and $\ep = \V \setminus \cp$, respectively. Then,
		\begin{itemize}
			\item[(i)] for any fixed $z \in \mc X_\ep$, the $\cp$-restricted game is potential;
			\item[(ii)] for any fixed $y \in \mc X_\cp$, the $\ep$-retricted game is potential.
		\end{itemize} 
		
	\end{lemma}
	
	The proof of Lemma \ref{lemma1}, which can be found in the Appendix, is based on an extension of the known result about the potential property of network coordination and anti-coordination games to the case of heterogeneous thresholds. More precisely, for a fixed configuration of the anti-coordinating players $z\in \mc X_\ep$, the $\cp$-restricted game is strategically equivalent to a network coordination game with heterogeneous thresholds $r_i^{(z)}$ for $i\in \cp$ that depend on both the threshold vector $\textbf{r}=(r_i)_{i\in \V}$ and the configuration of the anti-coordinating players $z$. The same reasoning applies to the $\ep$-restricted game. The claim then follows by observing that both the network coordination game and the network anti-coordination game with \textit{heterogeneous thresholds} are potential games. 

	As previously observed, even if the restrictions to players $\cp$ and $\ep$ are potential, the whole CAC game is not and the existence of Nash equilibria is not guaranteed. Under suitable assumptions we are able to assess the existence of Nash equilibria for the CAC game. Indeed, when the set of coordinating players is $\textbf{r}$-cohesive and they all play action $1$, coordinating players will be in equilibrium regardless of the actions of the anti-coordinating players. Then, the investigation on the existence of Nash equilibria reduces to the study of Nash equilibria of the $\mc S$-restricted game $u^{(\1)}$, that we proved to be a potential game. 
	Elaborating on this idea leads to following Theorem \ref{th:existence-extended}, that is an extension of Theorem 	\ref{coro:mixed-existence}.
	\begin{theorem}\label{th:existence-extended}
		Consider the mixed network CAC game on an undirected graph $\mc G=(\mc V,\mc E, W)$, with set of coordinating and anti-coordinating players $\cp\subseteq\mc V$ and $\ep=\mc V\setminus\cp$, respectively, and threshold vector $\mathbf{r}= (r_i)_{i \in \mc V}$. 
		\begin{enumerate}
			\item If $\cp$ is $\mathbf{r}$-cohesive then the game admits a pure strategy Nash equilibrium $x^* \in \{(\1,z):\,z\in \mc X_{\ep}\}$.
			\item  If $\cp$ is $(\1-\mathbf{r})$-cohesive then the game admits a pure strategy Nash equilibrium $x^* \in \{(\0,z):\,z\in \mc X_{\ep}\}$.
		\end{enumerate}
	\end{theorem}

		Theorem \ref{th:existence-extended} is a analogous to \cite[Corollary 3(i)]{Arditti.ea:2021}. Indeed, by setting $h_i = (1-2 r_i)w_i$, the network CAC game \eqref{eq:ut} is strategically equivalent to the mixed network coordination-anticoordination game model defined in \cite[Section VI]{Arditti.Como.Fagnani:2020}, up to a rescaling of the utilities of the latter by a factor $1/4$ and a relabelling of the actions. 

	The cohesiveness assumption guarantees the existence of Nash equilibria where coordinating players are at consensus, 
	but it proves insufficient to assure convergence to such set of the best response dynamics. As shown in the following proof to Theorem \ref{th:convergence-extended}, which extends Theorem \ref{coro:mixed-convergence}, convergence is obtain under the additional geometric assumption of indecomposability of the set of coordinating players in $\mc G$, which prevents the learning dynamics to be absorbed elsewhere. 
	Indeed, indecomposability guarantees uniqueness of the consensus configuration as Nash equilibrium of the $\cp$-restricted game for any $z\in \mc X_\ep $. This fact, combined with the potential property of the $\cp$-restricted game, guarantees the existence of a BR-path to a consensus configuration of the $\cp$ players from any initial condition. Once consensus is reached on the coordinating side, global reachability of a Nash equilibrium for the whole game is guaranteed by the potential property of the $\ep$-restricted game.
	\begin{theorem}\label{th:convergence-extended}
		Consider the network CAC game on an undirected graph $\mc G=(\mc V,\mc E, W)$, with set of coordinating and anti-coordinating players $\cp\subseteq\mc V$ and $\ep=\mc V\setminus\cp$ and threshold vector $\mathbf{r}= (r_i)_{i \in \mc V}$. 
		Assume that
		\begin{enumerate}
			\item[a)] $\cp$ is $\mathbf{r}$-cohesive or $(\1-\mathbf{r})$-cohesive,
			\item[b)] $\cp$ is $\mathbf{r}$-indecomposable in $\mc G$.
		\end{enumerate}
		Then 
		the set of Nash equilibria where coordinating players are at consensus is non-empty and it is globally reachable. 
	\end{theorem}
	
	 Theorem \ref{th:convergence-extended} should be compared to Corollary 3 (ii) in \cite{Arditti.ea:2021}. More precisely, Theorem \ref{th:convergence-extended} is a special case of \cite[Corollary 3 (ii)]{Arditti.ea:2021} up to the previously mentioned transformation that connects Theorem \ref{th:existence-extended} to \cite[Corollary 3(i)]{Arditti.ea:2021}. 
	
	\section{Conclusion}
	In this paper we have studied network games with heterogeneous players, some of them are playing a coordination game and some an anti-coordination game. We have established sufficient conditions in terms of the way coordinating players are located and wired that imply the existence of Nash equilibria that are consensuses on the side of the coordinating players. Under more restricted conditions we have also shown the convergence of best response learning rule to such equilibria.
	
	The techniques at the base of our results only rely on some features of the coordination and anti-coordination game. The key fact is that pure coordinating games are supermodular and that pure anti-coordination games are potential games.
	This paves the way to extend our results to wider families of games, including games defined on directed graphs, as in our recent work \cite{Arditti.ea:2021}. 
	\bibliographystyle{unsrt}
	\bibliography{bib}


\appendix
\textit{Notation for proofs.}It proves convenient to introduce the following notation. Given a set of players $\mc U\subseteq \mc V$, for $i \in \mc U$ and $y \in \mc X_{\mc U}$, let
	$$
	w_i^{\mc U,0}(y) = \sum_{j \in \mc U: y_j = 0} W_{ij}, \quad 	w_i^{\mc U,1}(y) = \sum_{j \in \mc U: y_j = 1} W_{ij}
	$$
	represent the restrictions of $i$'s degree 
	to players in $\mc U$ that play action $0$ and $1$, respectively, in a (restricted) configuration $y$.
	To lighten the notation, when $\mc U= \mc V$ we will simply denote 
	$w_i^0(x) \equiv w_i^{\mc V,0}(x)$ and $w_i^1(x) \equiv w_i^{\mc V,1}(x)$ for a player $i\in \mc V$ and a configuration $x\in \mc X$.

\begin{proof}[Lemma \ref{lemma1}]
	For every strategy profile $z$ in $\mc X_{\ep}$ of the anti-coordinating players, the utility function of the $\cp$-restricted game admits the following explicit expression
	\begin{equation}\label{eq:R-utlities}
		\begin{aligned}
			u_i^{(z)}(y) = &\sum_{j \in \cp} W_{ij} \left((1-r_i) y_iy_j + r_i (1-y_i)(1-y_j) \right) \\
			&+ y_i  \sum_{j \in \ep} W_{ij} \left( (1-r_i) z_j - r_i(1-z_j) \right)\\
			&+ \sum_{j \in \ep} W_{ij}r_i(1-z_j) \\
			= &\sum_{j \in \cp} W_{ij} \left((1-r_i^{(z)}) y_iy_j + r_i^{(z)} (1-y_i)(1-y_j) \right)\\
			&+ k_i^{(z)}(y_{-i})
		\end{aligned}
	\end{equation}
	where for every $i$ in $\cp$ and $y$ in $\mc X_{\cp}$
	$$
	r_i^{(z)} = r_i \left(1+ \frac{w_i^{\ep}}{w_i^{\cp}} \right) - \frac{w_i^{\ep,1}(z)}{w_i^{\cp}}
	$$
	and
	$$
	k_i^{(z)}(y_{-i}) = (r_i - r_i^{(z)}) w_i^{\cp,0}(y_{-i}) + r_iw_i^{\ep,0}(z).
	$$
	
	
	Let $\mc G_{\cp}=(\cp,\mc E_{\cp},W_{\cp,\cp})$ be the induced subgraph with  node set $\cp$, edge set $\mc E_{\cp}=\mc E\cap(\cp\times\cp)$ and weight matrix coinciding with the submatrix of $W$ obtained by keeping only rows and columns indexed both by elements of $\cp$. 
	Notice that the $\cp$-restricted game $u^{(z)}$ can be interpreted as a network coordination game on the induced subgraph $\mc G_{\cp}$ with modified thresholds $r_i^{(z)}$ and an additional non-strategic term $k_i^{(z)}$, which does not depend on the action of player $i \in \cp$.
	
	Since non-strategic terms do not affect the potentiality of games, it follows that the $\cp$ restricted game is potential with potential function
	\begin{equation}\label{eq:c-potential-nonuniform}
		\begin{aligned}
			\Phi^{(z)}_c(y) =&\frac{1}{2}\sum_{i,j\in \cp}W_{ij}\left(y_i y_j + (1-y_i)(1-y_j)\right)\\
			& - \sum_{i \in \cp}(r_i^{(z)}-\frac{1}{2})y_i w_i^{\cp}\,,
		\end{aligned}
	\end{equation}
	
	Analogously, for every $y$ in $\mc X_{\cp}$, the $\ep$-restricted game $u^{(y)}$ can be interpreted as network anti-coordination game on the induced subgraph $\mc G_{\ep}$ and by the same reasoning it follows that it is potential with potential function 
	\begin{equation}\label{eq:a-potential-nonuniform}
		\begin{aligned}
			\Phi^{(y)}_a(z) =&-\frac{1}{2}\sum_{i,j\in \ep}W_{ij}\left(z_i z_j + (1-z_i)(1-z_j)\right)\\
			& + \sum_{i \in \ep}(r_i^{(y)}-\frac{1}{2})z_i w_i^{\ep}\,,
		\end{aligned}
	\end{equation}
	where for every $i$ in $\ep$ the modified thresholds $r_i^{(y)}$ are defined as
	$$
	r_i^{(y)} = r_i \left(1+ \frac{w_i^{\cp}}{w_i^{\ep}} \right) - \frac{w_i^{\cp,1}(y)}{w_i^{\ep}}\,.
	$$
\end{proof}	
	\begin{proof}[Theorem \ref{th:existence-extended}]
		To prove the first statement, we exhibit a Nash equilibrium $x^*=(y^*, z^*)$, 
		where $y^* = \1$.
		First we show that the $\cp$-restricted game $u^{(z)}$ admits the equilibrium configuration $\1 \in \mc X_{\cp}$ for any $z \in \mc X_{\ep}$.
		Let $y^* = \1$. Setting $x = (y^*,z)$ with $z \in \mc X_{\ep}$, the best response of any $i\in \cp$ is $\{1\}$. Indeed, $1 \in \mc B_i(y^*,z)$ iff $w_i^1(x)\geq r_iw_i$ and we have
		$$
		w_i^1(x) = w_i^{\cp} + w_i^{\ep,1} \geq w_i^{\cp}  \geq  r_i w_i
		$$
		Notice that the last inequality holds true under the $\mathbf{r}$-cohesiveness assumption of $\cp$.
		This implies that $y^*$ is a Nash equilibrium for $u^{(z)}$ for any $z \in \mc X_{\ep}$.
		
		So we can focus on the $\ep$-restricted game $u^{(\1)}$. The game $u^{(\1)}$ is potential, so it admits at least a Nash equilibrium $z^*$.
		
		If we combine the two results we obtain that $x^*=(\1,z^*)$ is a Nash equilibrium, which concludes the proof. 
		
		By a similar reasoning we can prove statement 2 under the assumption of  $(\1-\mathbf{r})$-cohesiveness of $\cp$ and by setting $y^*=\0$.
	\end{proof}
	\begin{proof}[Theorem \ref{th:convergence-extended}]
		We show that for each configuration $x \in \mc X$ there exists a BR-path $x \to x^*$ where $x^*=(y^*,z^*)$ is a Nash equilibrium and $y^* \in \{ \0_{\cp}, \1_{\cp}\}$.
		We construct the BR-path $x \to x^*$ by a two-step procedure.
		\begin{itemize}
			\item Step I: we construct a BR-path $x \to (y^*, z)$ such that only players in $\cp$ update their actions.
			
			\item Step II: we construct a BR-path $(y^*, z) \to x^*$ such that only players in $\ep$ update their actions.
		\end{itemize}
		
		\textbf{Step I}. Consider the $\cp$-restricted 
		game $u^{(z)}$, where players in $\ep$ are in fixed in state $z \in \mc X_\ep$.

		We begin by proving that the Nash equilibria of $u^{(z)}$ form a non-empty subset of $\{\1,\0\}$.
		
		First we show that the set of Nash equilibria is non-empty.
		Under the $\mathbf{r}$-cohesiveness assumption of $\cp$, similarily to what done in the proof of Theorem \ref{th:existence-extended}, we can set $y^*=\1$ and show that $1 \in \mc B_i(y^*,z)$, $\forall i \in \cp$. Indeed $\forall i \in \cp$ it holds that
		$$
		w_i^1(x) = w_i^{\cp} + w_i^{\ep,1} \geq w_i^{\cp}  \geq  r_i w_i .
		$$
		Similarly, under the $(\1-\mathbf{r})$-cohesiveness assumption we can set $y^*=0 \1$ and show that $0 \in \mc B_i(y^*,z)$, $\forall i \in \cp$, as
		$$
		w_i^0(x) = w_i^{\cp} + w_i^{\ep,0} \geq w_i^{\cp}  \geq  (1-r_i) w_i .
		$$

		In order to prove that  $u^{(z)}$ does not possess Nash equilibria outside of $\{\1,\0\}$, we consider an action configuration $y$ such that $y\notin \{\1, \0\}$ and we show that it is not a Nash equilibrium of $u^{(z)}$. 
		Consider the following partition of $\cp$:
		$$
		\begin{aligned}
			\cp_1 &= \left\lbrace i \in \cp : y_i = 1 \right\rbrace \neq \emptyset, \\
			\cp_0 &=  \left\lbrace i \in \cp : y_i = 0 \right\rbrace \neq \emptyset.
		\end{aligned}
		$$
		Then, by assumption b), we know that $\exists i \in \cp$ such that one of the two equations \eqref{eq:indec1} or \eqref{eq:indec0} holds true.
		
		We first consider the case where \eqref{eq:indec1} is satisfied.
		Recall that, if $i\in \cp_1$, then $y_i = 1$. In such case we have that $1 \notin \mc B_i(y,z)$, since
		$$
		w_i^{\cp_1} + w_i^{\ep,1} < w_i^{\cp_1} + w_i^{\ep} < r_i w_i
		$$
		where the last inequality follows from \eqref{eq:indec1}.
		Then we have shown that there exists a player $i \in \cp$ who can improve her utility by changing her action in configuration $y$. As a result, $y$ is not a Nash equilibrium for $u^{(z)}$.
		
		If \eqref{eq:indec0} is satisfied, we can show that $y$ is not a Nash equilibrium for $u^{(z)}$ by following the exact same reasoning of the previous case.
		
		Recall that, for any fixed $z\in \mc X_{\ep}$, the $\cp$-restricted 
		game $u^{(z)}$ is potential. This property guarantees that for any initial condition $y \in \mc X_{\cp}$, there exists a BR-path $y \to y^*$ where $y^*$ is a Nash equilibrium of $u^{(z)}$. By what already shown, $y^*$ is  necessarily either $\1$ or $\0$. We extend the path $y \to y^*$ to players in $\ep$ by setting their actions to $z$, thus obtaining the BR-path
		$$
		x \to (y^*, z).
		$$
		
		\textbf{Step II}. Consider the $\ep$-restricted network anti-coordination game $u^{(y^*)}$, where the actions of the coordinating players are fixed to $y^* \in \{\1, \0 \}$. The game $u^{(y^*)}$ is potential, so there exists a BR-path 
		$$
		z \to z^*
		$$
		where $z^*$ is a Nash equilibrium of $u^{(y^*)}$. We extend this path to players in $\cp$ by setting their actions to $y^*$ thus obtaining the BR-path
		$$
		(y^*,z) \to (y^*,z^*) = x^* .
		$$
		
		If $y^*$ is a Nash equilibrium for $u^{(z^*)}$, then $x^*$ is a Nash equilibrium for the whole game as
		\begin{itemize}
			\item $\forall j \in \cp$,  $x^*_j \in \mc B_j(x^*)$ since $y^*$ is a Nash equilibrium of $u^{(z^*)}$,
			\item $\forall j \in \ep$, $x^*_j \in \mc B_j(x^*)$ since $z^*$ is a Nash equilibrium of the game $u^{(y^*)}$,
		\end{itemize}
		and the proof is complete.
		If instead $y^*$ is not a Nash equilibrium for $u^{(z^*)}$, it must be the case that either $\cp$ is  $\mathbf{r}$-cohesive and $y^*=\0$ or $\cp$ is $(\1-\mathbf{r})$-cohesive and $y^*=\1$. In both cases $\1 - y^*$ is the only Nash equilibria of $u^{(z^*)}$ and as in step 1 we can construct a BR-path $y^* \to y^{**}=\1-y^*$. This path can be extended to players in $\ep$ by setting their actions to $z^*$ thus obtaining the BR-path
		$$
		x^* = (y^*,z^*) \to (y^{**}, z^*) .
		$$
		We then proceed as in step 2. to find a configuration $z^{**}$ which is a Nash equilibrium of $u^{(y^{**})}$ and a BR-path
		$$
		(y^{**},z^*) \to (y^{**},z^{**}) = x^{**} .
		$$
		Notice that it is guaranteed that $y^{**}$ is a Nash equilibrium of $u^{(z^{**})}$ because either $\cp$ is $r$-cohesive and $y^{**}=\1$ or $\cp$ is $(1-r)$-cohesive and $y^{**}=\0$ so that $y^{**}$ is a Nash equilibrium of $u^{(z)}$ for any $z \in \mc X_\ep$.
		This concludes the proof.
\end{proof}
\end{document}